\documentclass[12pt]{iopart}

\usepackage{iopams} 
\usepackage{graphicx}
\usepackage{url}

\newcommand{\sech}{\mathrm{sech}}

\begin{document}

\title{Pulsed source of spectrally uncorrelated and indistinguishable photons at telecom wavelengths.}

\author{N.~Bruno, A.~Martin, T.~Guerreiro, B.~Sanguinetti and R.~T.~Thew}

\address{Group of Applied Physics, University of Geneva, Switzerland}
\ead{robert.thew@unige.ch}

\begin{abstract}

We report on the generation of indistinguishable photon pairs at telecom wavelengths based on a type-II parametric down conversion process in a periodically poled potassium titanyl phosphate (PPKTP) crystal. The phase matching, pump laser characteristics and coupling geometry are optimised to obtain spectrally uncorrelated photons with high coupling efficiencies. Four photons are generated by a counter-propagating pump in the same crystal and anlysed via two photon interference experiments between photons from each pair source as well as joint spectral  and g$^{(2)}$ measurements. We obtain a spectral purity of 0.91 and coupling efficiencies around 90\% for all four photons without any filtering. These pure indistinguishable photon sources at telecom wavelengths are perfectly adapted for quantum network demonstrations and other multi-photon protocols.

\end{abstract}

\maketitle
\tableofcontents
\section{Introduction}

Engineering multi-photon quantum states of light is becoming increasingly important as we move beyond point-to-point, or two-party systems, entangled systems and towards entanglement based protocols such as quantum teleportation, swapping and relays, useful for QKD~\cite{Gisin2007}. Already entanglement swapping protocols rely on the interference between quantum states carried by independent photons~\cite{Riedmatten2005,Kaltenbaek2009}. As we look towards more complex quantum networks~\cite{Kimble2008} and simulation experiments~\cite{Aspuru-Guzik2012}, efficient coupling and spectrally pure photon generation is of critical importance. 

Currently, spontaneous parametric down conversion (SPDC) is the simplest and most versatile way to generate photon pairs and historically PPLN waveguides have been exploited in the telecom regime due to their high brightness~\cite{Tanzilli2012}. However, most of the time the twin photons generated in this way have spectral correlations, which decrease the purity of their quantum state~\cite{Grice2001}. Spectral filtering is usually employed to erase these correlations, consequently introducing additional losses~\cite{Halder2007,Fulconis2007,Martin2012b,Bruno2013}. Another solution is to adapt the pump laser spectrum and the dispersion properties of the non-linear material. This approach has been implemented in various media such as bulk crystals~\cite{Mosley2008,Mosley_conditional_2008}, waveguides~\cite{Spring2013,Harder2013}, or fibers~\cite{Soller2011}. Quantum communication requires narrowband sources and recent focus has shifted to using periodically poled potassium titanyl phosphate (PPKTP) nonlinear crystals with ps pulsed lasers~\cite{Evans2010,Yabuno2012,Jin2013,Ikuta2013} to address these demands of quantum communication. We have combined this approach with some of our recent work on optimising source-to-fibre coupling~\cite{Guerreiro2013} to realise multi-photon experiments with pure, indistinguishable narrowband telecom photons with high levels of purity and coupling efficiencies that  are ideally suited to testing quantum network protocols.

After a presentation of the theoretical model employed to simulate the non-linear interaction in the crystal, we describe the experimental setup. In the third part, we present the characterization of the purity of the photons emitted by our source. In the last part, we report the brightness of the source.

\section{Theoretical Model for Phase Matching and Photon Coupling}
In the following we outline the theory used for modeling the phase matching as well as the focusing characteristics for the pump laser and photon pairs that are needed to optimise both the purity and the fibre coupling. SPDC processes are governed by energy and momentum conservation laws: 
\begin{equation}
\omega_p = \omega_i + \omega_s ; \qquad
\vec k_p = \vec k_i + \vec k_s + \frac{2 \pi}{\Lambda} \vec z
\end{equation}
where $\omega$ and $\vec k$ represent the frequencies and the wavevectors of the pump (p), signal (s), and idler (i) photons, respectively. In a periodically poled crystal, a poling period $\Lambda$ is inscribed in the crystal to compensate the dispersion and achieve quasi-phase matching~\cite{Fejer1992}. The non-linear process that takes place in this kind of crystal is described by a Hamiltonian function of the creation and annihilation operators of the signal and idler fields, of the following form:
\begin{equation}\label{eq_nl_int}
H = c\int d\omega_s  \, d\omega_i \, d\vec k_s \, d\vec k_i \, d\vec k_p \, S(\omega_s,\omega_i,\vec k_s,\vec k_i,\vec k_p) \, a^\dag(\omega_s,\vec k_s) \, a^\dag(\omega_i,\vec k_i) + \rm h.c. .
\end{equation}
We define the joint spectral amplitude as $S(\omega_s,\omega_i,\vec k_s,\vec k_i,\vec k_p)=\epsilon(\omega_s,\omega_i)\phi(\omega_s,\omega_i,\vec k_s,\vec k_i,\vec k_p)$, where $\epsilon(\omega_s,\omega_i)$ and $\phi(\omega_s,\omega_i,\vec k_s,\vec k_i,\vec k_p)$ are the pump energy envelope and the phase matching function, respectively. These two functions depend on the energy conservation and the phase matching conditions. For a Sech$^2$-shaped pump pulse, such as a pulse generated in a mode-locked laser, the first factor is given by:
\begin{equation}\label{eq_pompe_shape}
\epsilon(\omega_s,\omega_i) \propto \sech \left[ \frac{(\omega_i+\omega_s-\omega_p)}{3\pi } \cdot \frac{2 \log(2+\sqrt{3})}{\Delta \omega_p}\right] 
\end{equation}
 with $\Delta \omega_p$ the full width at half maximum of the pump pulse in frequency. In the Gaussian approximation, this becomes:
\begin{equation}
 \epsilon(\omega_s,\omega_i) \propto  \exp{\left( \frac{\omega_i+\omega_s-\omega_p}{4 \sigma_p^2}\right) }
\end{equation}  
with $\sigma_p = \Delta \omega_p/2\sqrt{2 ln 2} $.
 The phase matching function for a crystal of a length $L$ is given by: 
 \begin{equation}\label{eq_phase_matching}
 \phi (\omega_s,\omega_i,\hat s_s,\hat s_i,\hat s_p) \propto  {\rm sinc}\left[   \frac{L}{2}\left(  k_i\left( \omega_i\right)  \hat s_i+ k_s\left(  \omega_s\right)  \hat s_s + \frac{2 \pi}{\Lambda} \hat z- k_p\left( \omega_p\right)  \hat s_p\right)  \right]    ,
 \end{equation}
 with $\hat s_j$ the unit vector of $\vec k_j$ and $\hat z$ the direction of the poling\cite{Ljunggren2005}.
Considering only the wave vector along the $\hat z$ axis and again the Gaussian approximation this function can be written as:
\begin{equation}
\phi(\omega_s,\omega_i) \propto  \exp{\left( -\alpha^2 \frac{\Delta k^2 L^2}{4}\right) } ,
\end{equation}\label{eq:gaussian}
with $\alpha = 0.439$ and $\Delta k = k_i(\omega_i)+ k_s(\omega_s) + \frac{2 \pi}{\Lambda} - k_p(\omega_p)$. Furthermore, for very small deviations from the central wavelengths of the pump  $\omega_{p,0}$ and of the emitted photons $\omega_{i,0}$ and $\omega_{s,0}$, we have $\Delta k = (\omega_i-\omega_{i,0})k'_i+ (\omega_s-\omega_{s,0})k'_s - (\omega_p-\omega_{p,0})k'_p$, with $k'_j = \left. \frac{\partial k_j(\omega_j)}{\partial \omega_j}\right|_{\omega_{j,0}}$. 
By using these approximations, the joint spectral amplitude can be rewritten as a Gaussian function~ \cite{Grice2001,Osorio2013}:
\begin{eqnarray}\label{eq_JSA}
S(\Delta \omega_s, \Delta \omega_i) & = & \epsilon(\omega_s,\omega_i)\varphi(\omega_s,\omega_i) \\
& \propto & {\rm exp} \left[-\left( \frac{1}{\sigma_p^2} + \alpha^2L^2(k'_p- k'_s)^2\right) \frac{\Delta \omega_s^2}{4}  \right. \\
& & \left. -\left( \frac{1}{\sigma_p^2} + \alpha^2L^2(k'_p- k'_i)^2\right)  \frac{\Delta \omega_i^2}{4}  \right.  \\
&  &\left. -\left( \frac{1}{\sigma_p^2} + \alpha^2L^2(k'_p- k'_s)(k'_p- k'_i)\right)  \frac{\Delta \omega_s\Delta \omega_i}{2} \right],
\end{eqnarray}
with $ \Delta \omega_j = \omega_j - \omega_{j,0}$.
The purity of the two-photons state is defined as the inverse of the number of Schmidt modes $\mathcal{K}$, more precisely $\mathcal{P}=1/\mathcal{K}={\rm Tr}[{\rm Tr}_i[\rho]]$.
To generate pairs of photons in a separable state, i. e. $\mathcal{K} = 1$, it must be:
\begin{equation}\label{eq_purity_cond}
\alpha^2L^2\sigma_p^2(k_p'-k_s')(k_p'-k_i') = -1.
\end{equation}
To satisfy this equation, the condition $k_s'<k_p'<k_i'$ must be fulfilled in the non-linear medium. 

Following equation (\ref{eq_purity_cond}), we find that a PPKTP crystal can satisfy these constraints. To satisfy the phase matching condition, a poling period of 47.8\,$\mu$m inscribed in the crystal enables the type-II non-linear interaction between pump, signal and idler photons polarised along the X, Z, and X axes of the crystal, respectively. In this configuration, we obtain $(k_p'-k_s')(k_p'-k_i')= 2.63\times 10^{-2}$\,m$^{-1}$.GHz$^{-1}$. Thus, for a 3\,cm long crystal a purity close to one should be achievable with a pump bandwidth of 467\,rad$^{-1}$\,GHz, which corresponds to a pulse duration with FWHM = $2.6$\,ps, assuming a Gaussian approximation of the pump as in equation~(\ref{eq:gaussian}). In this way we are able to generate degenerate photon pairs at 1544\,nm with a pump laser at 772\,nm.

If we consider these experimental characteristics, we can simulate the non-linear interaction defined in equation (\ref{eq_nl_int}) without any approximations, \textit{i.e.} taking into account the shape of the pump, equation~(\ref{eq_pompe_shape}), the crystal geometry and the wavevector orientations of equation~(\ref{eq_phase_matching}). This simulation gives us access to all of the correlations for the signal and idler photon in wavelength, wavevector, as well as joint wavelength - wavevector basis~\cite{Guerreiro2013}. The first is useful in order to estimate the purity in frequency, and the other two are used to determine the coupling efficiencies of the generated twin photons into single mode fibres. To obtain pure photons, we need to reduce the correlation between the signal and idler wavelengths. As mentioned above, the wavelength correlation depends on the type and length of the crystal, on the pulse duration of the pump laser, and also on the spatial profile of the three fields. To have an efficient coupling, one needs to reduce the wavelength - wavevector correlations for the signal and idler photons and increase the signal-idler wavevector correlation (see Ref.~\cite{Guerreiro2013} for more detail). Thus, from this simulation we can study the influence of the spatial modes of the fields on the spectral purity and on the coupling efficiencies. 

\begin{figure}[h]
\center
\includegraphics[width=0.6\linewidth]{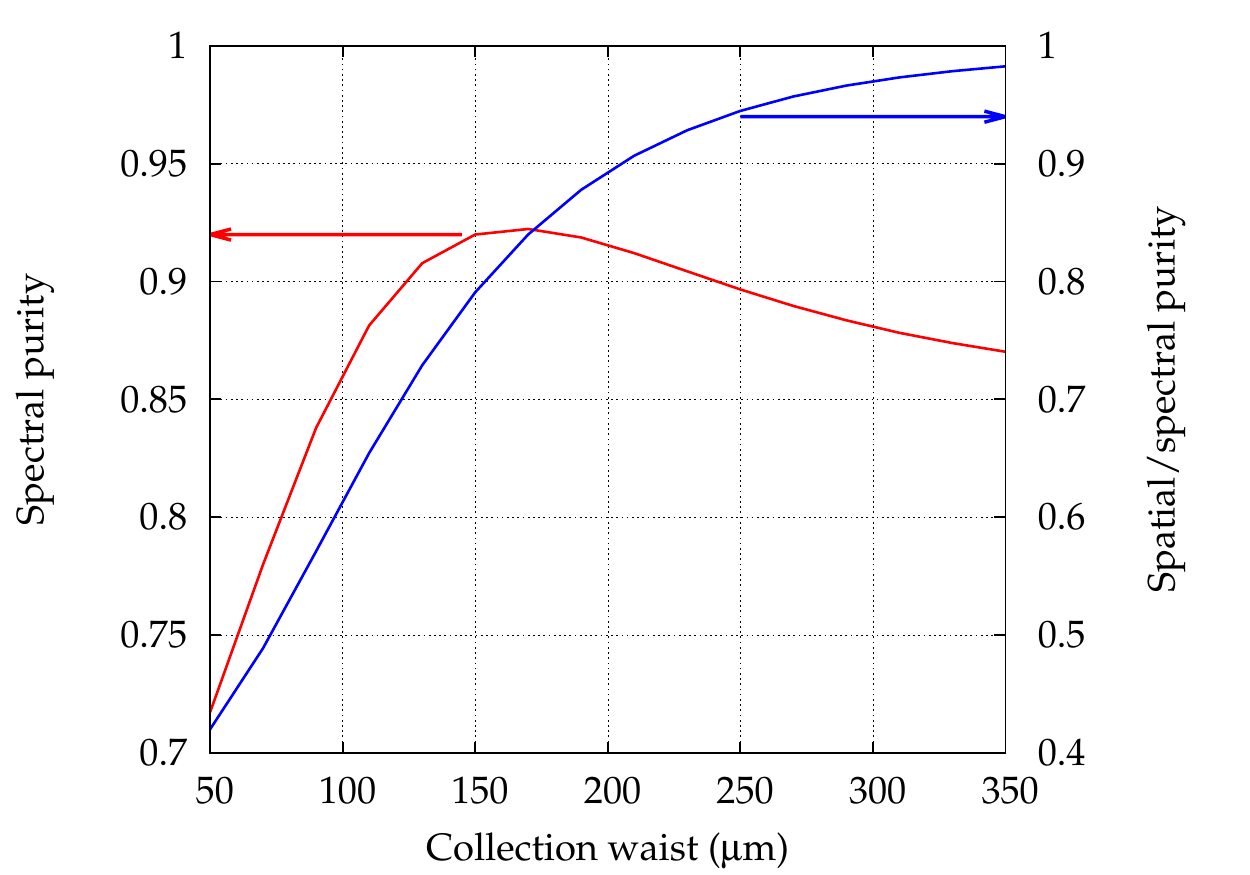}
\caption{\label{fig_purity}Numerical simulation of the photon pair spectral purity (red curve) and signal spatial-spectral purity (blue curve) for a 3\,cm long PPKTP bulk crystal pumped by a 772\,nm pump laser with a bandwidth of 467\,rad$^{-1}$\,GHz.}
\end{figure}

The wavevector correlations only depend on the wavevector of the pump, and to be maximised, the pump field should be close to a plane wave. Consequently, the focusing parameter denoted $\xi$, defined as the ratio between the half length of the crystal and the Rayleigh range of the beam $z_R$ ($\xi = L/2z_R$), needs to be close to zero. For this reason we choose a waist of 296\,$\mu$m for the pump laser in the center of the crystal, which corresponds to a focal parameter of 0.0425. \figurename{~\ref{fig_purity}} illustrates the dependence on purity and coupling.  To define the optimal focal parameter for the photon pairs generated in the crystal, we studied, as shown in~\figurename{~\ref{fig_purity}}, the photon pair spectral purity and the photon's spatial-spectral purity, as a function of the photon collection waist. We chose a collection waist of 187\,$\mu$m ($\xi = 0.212$), which is a good compromise between spectral purity and coupling efficiency. To obtain this waist, lenses with a 400\,mm focal length need to be placed at the output of the crystal combined with 11\,mm focal length lenses to then couple into the single mode fibres.

\section{Experimental realisation}

\begin{figure}[h]
\centering
\includegraphics[width=0.8\linewidth]{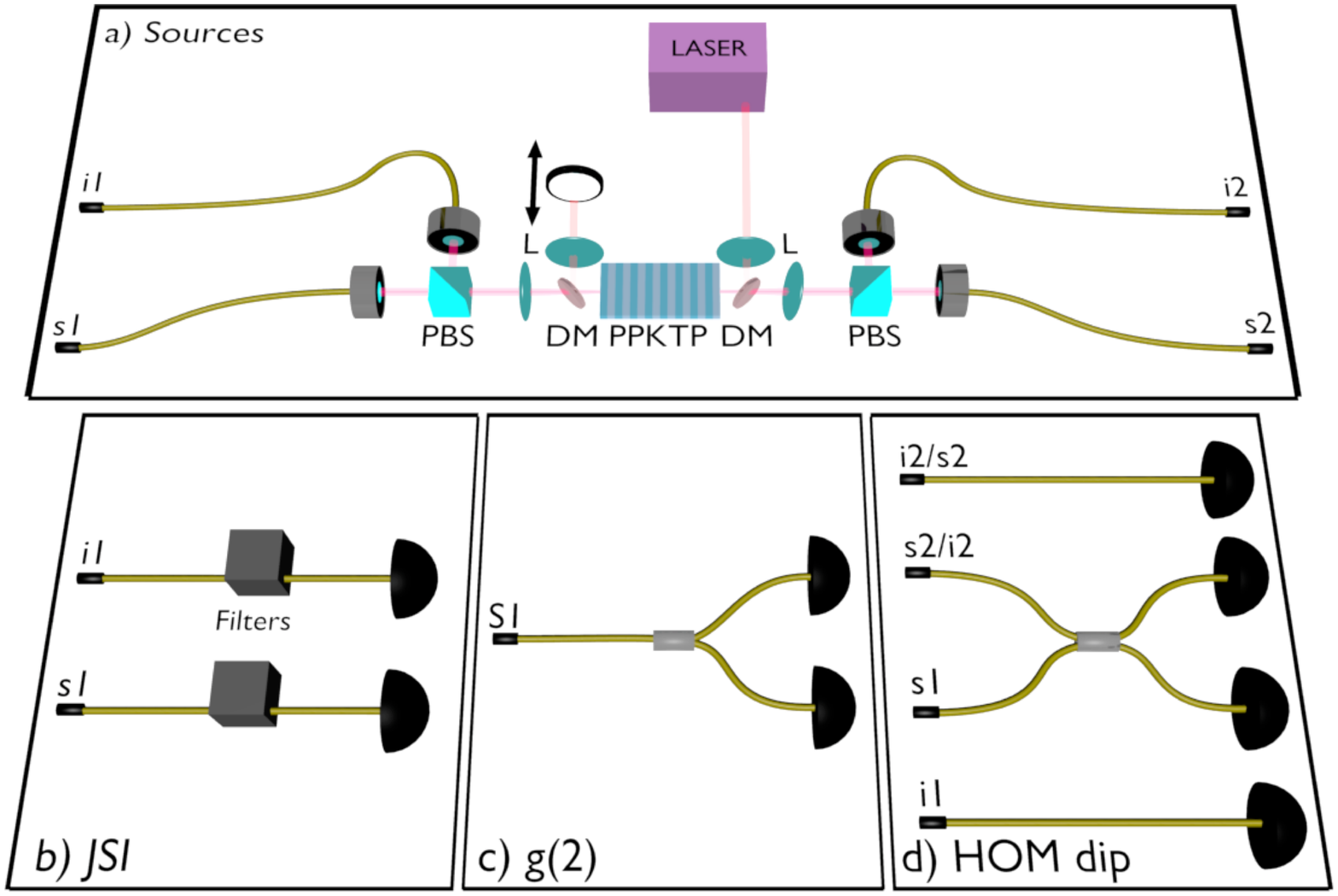}
\caption{\label{fig_setup} a) Experimental setup. A 772nm ps-pulsed laser pumps a periodically poled potassium titanylphosphate (PPKTP) crystal in a double-pass configuration. Dichroic mirrors (D) separate the telecom and visible wavelengths. A variable delay is placed before the second passage of the pump laser through the crystal. The generated photon pairs are first collimated by a set of lenses (L) and then separated with polarising beam-splitters (PBS) and coupled into single mode fibres. Three characterisation techniques are shown: b) the joint spectral intensity; c) the second order autocorrelation function $g^2(\tau)$, and d) Hong-Ou-Mandel (HOM) interference.}
\label{source}
\end{figure}

A schematic of the experimental setup is given in~\figurename{~\ref{source}}. The light from a Ti-Sapphire picosecond mode-locked laser at 772\,nm is injected into a 3~cm long PPKTP bulk crystal and we exploit a double-pass configuration.
The source is built in a double-pass configuation via a set of two dichroic mirrors employed to inject the laser and separate it from the generated photons. This enables the production of two photon pairs emitted independently in opposite directions, with an adjustable delay.
The photon pairs are deterministically separated and all four photons are then coupled into single mode telecom fibres. All four photons are generated around 1544\,nm and we measure a fibre coupling efficiency of $(90 \pm  4) \%$ for all four photons.

\section{Characterisation of the spectral purity}

There are several different techniques for characterising the spectral purity of the photons. We test three of these, which are represented in the lower half of~\figurename{~\ref{source}}: the second order autocorrelation function ($g^{(2)}(\tau)$) for each photon, the joint spectral intensity for the pair of photons from each source, and finally the two-photon (HOM) interference between photons from independent sources.

\begin{figure}[h]
\center
\includegraphics[width=0.6\linewidth]{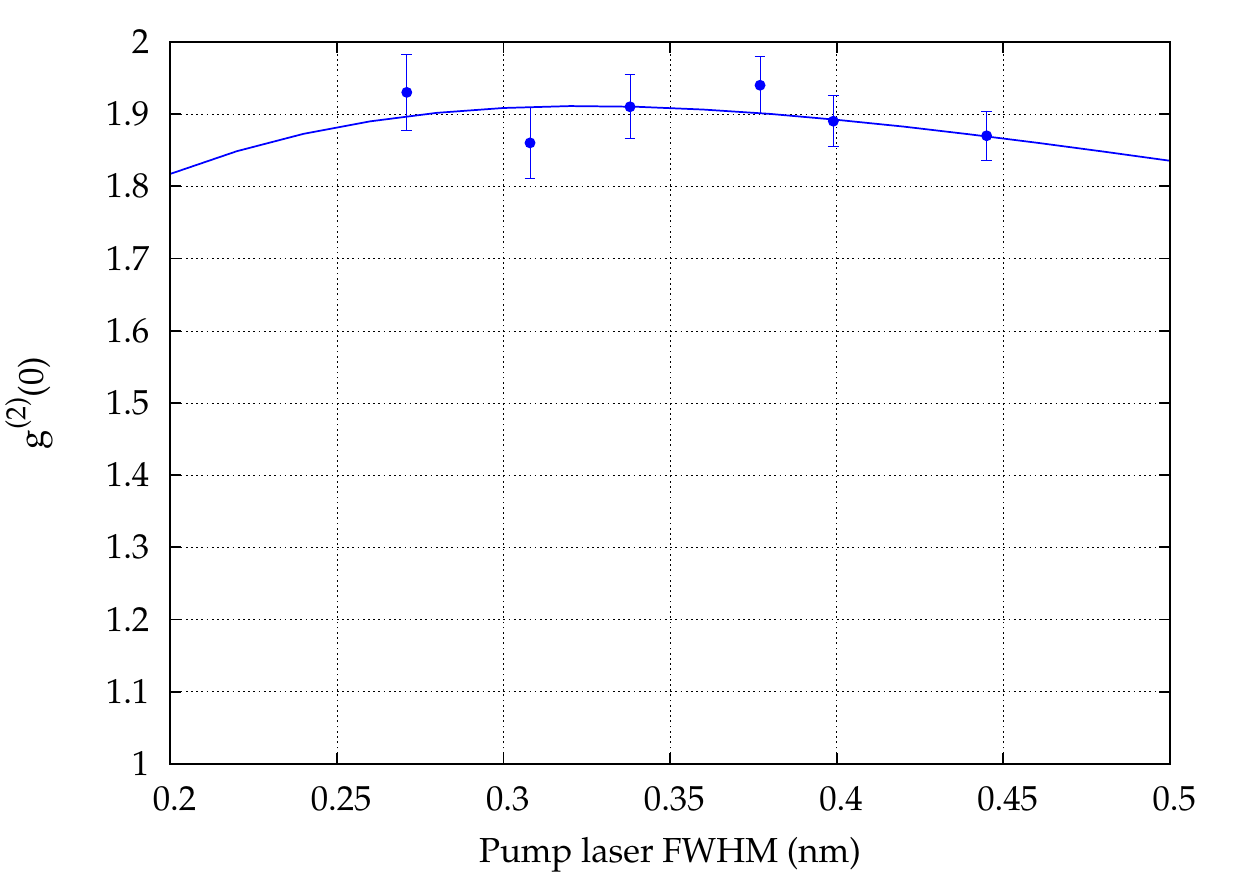}
\caption{\label{fig_g2}$g^{(2)}(0)$ as a function of the laser pump FWHM. The point and the line represent the experimental data and the simulation predications, respectively.}
\end{figure}

The second order autocorrelation function ($g^{(2)}(\tau)$) in a Hanbury Brown and Twiss like experiment~\cite{BROWN1956}, involves sending the signal photons to a balanced beam splitter and placing two single photon detectors based on InGaAs avalanche photodiodes (APD) at the output. A time-to-digital convertor (TDC) is used to record the coincidence histogram as a function of the time difference between clicks on the two detectors. This measurement allows one to estimate the photon number statistics.  For a pure photon we expect a $g^{(2)}(0)$ close to 2, which corresponds to thermal statistics typical of a single mode, and when the number of modes increases $g^{(2)}(0)$ goes to 1 which is the signature of a Poissonian distribution, characteristic of a multimode emission from the nonlinear medium~\cite{Tapster1998}. For a SPDC source we can define the $g^{(2)}(0) = 1+\mathcal{P} = 1+ \frac{1}{\mathcal{K}}$, where $\mathcal{P}$ and $\mathcal{K}$ represents the purity and the number of Schmidt modes, respectively~\cite{Christ2011,Sekatski2012,Bruno2013}. The measurement of $g^{(2)}(0)$ as a function of the pump laser bandwidth is reported in \figurename{~\ref{fig_g2}}. The maximum of purity is obtained for a pump laser with a FWHM equal to 0.33\,nm as predicted by the numerical simulation. For the following measurements, we use this value. This results in an experimentally measured value of  $g^{(2)}(0) = 1.91 \pm 0.04$, corresponding to a purity of $\mathcal{P} = 91\%\pm 4\%$.

Another quantity that is useful in order to evaluate the purity of the two photon state is the joint spectral intensity (JSI). To measure it we use two tunable filters with a bandwidth of 0.2\,nm, one for each photon of the independent pairs. Two InGaAs APDs are placed at the output of the filters and connected to a coincidence measurement apparatus. By recording the coincidence rate as a function of the position of the two filters, the JSI can be reconstructed.
\begin{figure}[h!]
\begin{tabular}{cc}
a)&b)\\
\includegraphics[width=0.45\linewidth]{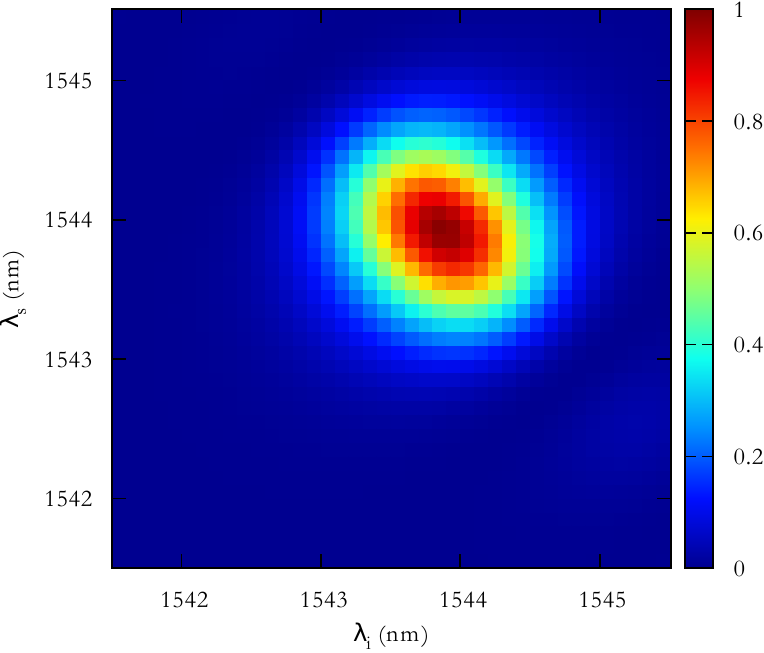}&
\includegraphics[width=0.465\linewidth]{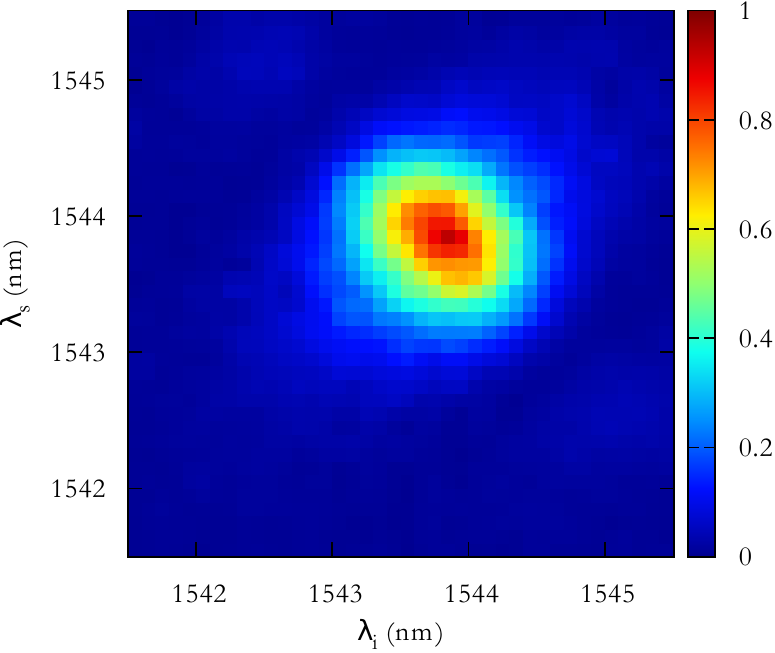}
\end{tabular}
\caption{\label{fig_JSI} Joint spectral intensity. The simulation (a) and experimental results (b) for the configuration described in the text.}
\end{figure}

 \figurename{~\ref{fig_JSI} a)} and b) represent the simulated and measured JSIs, respectively, and it shows the good agreement between the theoretical prediction and the experimental results. The simulation predicts a purity of 91.8\%, and we obtain, by fitting the experimental data with a two dimensional gaussian function defined in equation (\ref{eq_JSA}), a purity of $91\% \pm 3\%$. N. b. that the sidepeaks typical of a squared $sinc$ function are attenuated as an effect of the spatial filtering with single mode fibres, given that the collection waist is optimised to reduce wavelength-wavevector correlations and at the same time improve the spectral purity.

To validate this result and prove both the indistinguishability and purity of the four photons generated in our double pass source, we performed a series of two photon interference experiments. For each source, each direction of emission, one photon of the generated pair is directly sent to an InGaAs APD. Thus, when the two detectors fire they herald the creation of two photon pairs emitted in opposite directions. The other two photons are sent to a fibre 50/50 beamsplitter (BS). Another two APDs are placed at the output of the BS. The four detectors are linked to a time-to-digital convertor (TDC), which records time stamps for the four-fold coincidences as a function of the delay between the arrival times of the two photons that interfere at the BS. The delay is varied via a motorised linear translation stage attached to the mirror which reflects the pump laser before the second passage through the crystal (see \figurename{~\ref{fig_setup}).

\begin{figure}[h]
\begin{tabular}{cc}
a)&b)\\
\includegraphics[width=0.5\linewidth]{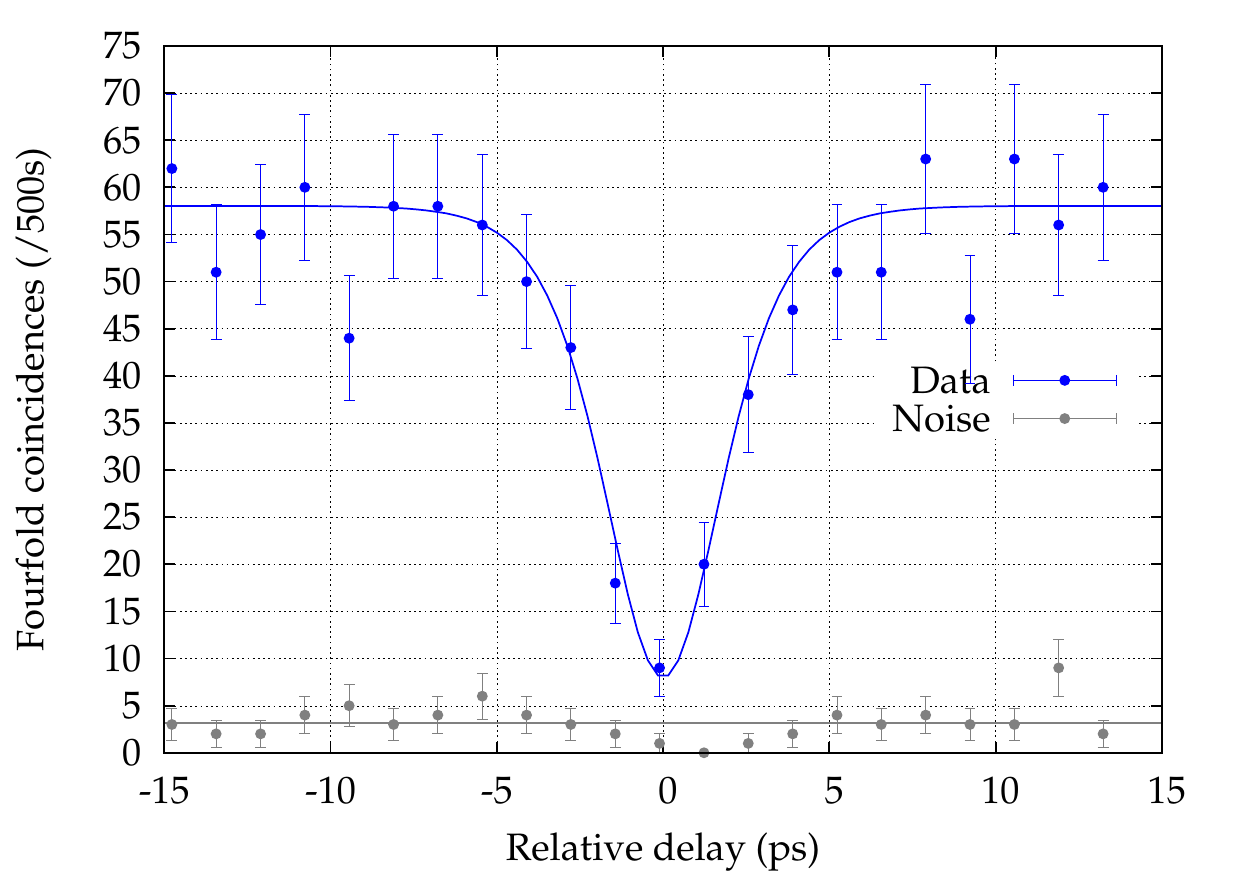}&
\includegraphics[width=0.5\linewidth]{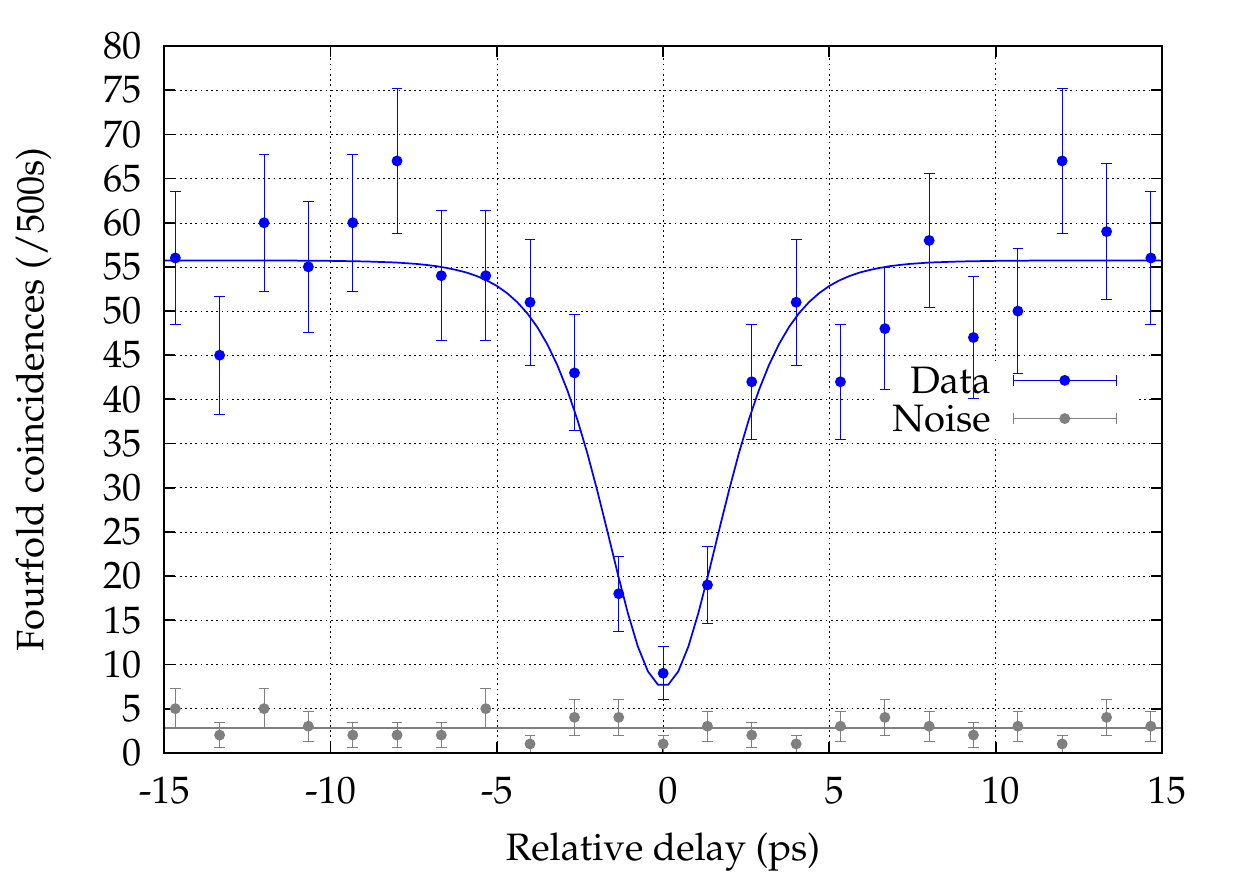}
\end{tabular}
\caption{\label{fig_dip} HOM dips obtained between signal-signal (a) and signal-idler photons from independent sources as a function of the delay between the photons. The blue and black points represent the four-fold coincidences and accidental coincidences, respectively. For both configurations a visibility of $91\pm2$\% is obtained.}
\end{figure}

The results shown in \figurename{~\ref{fig_dip}} a) and b) represent the Hong-Ou-Mandel (HOM) dips between the signal from one source and the signal and idler photons from the other source, respectively. In both configurations we obtain net visibilities of $91\pm 2$\%. The interference visibility is given by :
\begin{equation}
\rm V = Tr[\rho_a\rho_b] = \frac{Tr[\rho_a^2]+Tr[\rho_b^2]-||\rho_a-\rho_b||}{2},
\end{equation}
where $\rm \rho_a$ and $\rm \rho_b$ represent the density matrices of the two photons at the input of the beam splitter. Accordingly, the visibility measured for the signal-signal photons interference directly gives the spectral purity, while for the dip between a signal and an idler photon the visibility gives also information about the distinguishability. The obtained visibilities prove that less then 1.1 spectral modes are emitted by the crystal and that the four photons are perfectly spectrally indistinguishable, as predicted by the numerical simulation.

\section{Source brightness}

As well as the photon purity, one of the key features of this source is its brightness, since it provides rates that allow one to concretely implement quantum communication tasks. We define the brightness as the number of photon pairs available at the output of the source per mode and pump power. The brightness depends on the pair creation efficiency and on the transmission losses of the source. For a pulsed laser with pump power of 200\,mW, the crystal generates 0.01 photons per pulse at a repetition rate of 80\,MHz, which correspond to 4.0 pairs/$\mu$J. At the fibred output of the source we obtain a brightness of 2.6 pairs/$\mu$J, the losses per photon being of 0.96\,dB, 0.50\,dB due to the optical elements which build the source and 0.46\,dB introduced by the coupling into single mode fibre. It can be noted that for all the presented results the laser pump power is set to 600\,pJ per pulse to generate just 0.0025 photon pairs in order to avoid errors due to double pair contributions.

\section{Conclusion}

We have presented an experimental set-up for generating pure indistinguishable narrowband telecom photons in a double-pass configuration using a PPKTP crystal. By combining concepts for pure photon generation based on phase matching non-linear crystals and pump laser bandwidths along with the constraints of coupling photons into single mode fibres we obtain high coupling efficiencies and purities of around 90\,\% for all four photons. The type II phase matching condition allows us to deterministically separate all four photons providing us with a flexible multi-photon source of photons for diverse applications in quantum communication and networking.

\ack
The authors thank H. Zbinden, N. Gisin, C. Osorio and J. Schaake for useful discussions. This work was partially supported by the Swiss  NCCR-QSIT project.

\section*{References}
\bibliographystyle{iopart-num}
\bibliography{bib}

\providecommand{\newblock}{}
\begin{thebibliography}{10}
\expandafter\ifx\csname url\endcsname\relax
  \def\url#1{{\tt #1}}\fi
\expandafter\ifx\csname urlprefix\endcsname\relax\def\urlprefix{URL }\fi
\providecommand{\eprint}[2][]{\url{#2}}

\bibitem{Gisin2007}
Gisin N and Thew R 2007 {\em Nat. Photonics\/} {\bf 1} 165

\bibitem{Riedmatten2005}
de~Riedmatten H, Marcikic I, van Houwelingen J~A~W, Tittel W, Zbinden H and
  Gisin N 2005 {\em Phys. Rev. A\/} {\bf 71} 050302

\bibitem{Kaltenbaek2009}
Kaltenbaek R, Prevedel R, Aspelmeyer M and Zeilinger A 2009 {\em Phys. Rev.
  A\/} {\bf 79} 040302
  \urlprefix\url{http://link.aps.org/doi/10.1103/PhysRevA.79.040302}

\bibitem{Kimble2008}
Kimble H~J 2008 {\em Nature\/} {\bf 453} 1023--1030

\bibitem{Aspuru-Guzik2012}
Aspuru-Guzik A and Walther P 2012 {\em Nat. Phys.\/} {\bf 8} 285--291
  \urlprefix\url{http://www.nature.com/doifinder/10.1038/nphys2253}

\bibitem{Tanzilli2012}
Tanzilli S, Martin A, Kaiser F, {De Micheli} M~P, Alibart O and Ostrowsky D~B
  2012 {\em Laser Photonics Rev.\/} {\bf 6} 115--143
  \urlprefix\url{http://doi.wiley.com/10.1002/lpor.201100010}

\bibitem{Grice2001}
Grice W, U'ren A and Walmsley I 2001 {\em Phys. Rev. A\/} {\bf 64} 063815

\bibitem{Halder2007}
Halder M, Beveratos A, Gisin N, Scarani V, Simon C and Zbinden H 2007 {\em Nat.
  Phys.\/} {\bf 3} 692--695
  \urlprefix\url{http://www.nature.com/doifinder/10.1038/nphys700}

\bibitem{Fulconis2007}
Fulconis J, Alibart O, O'Brien J~L, Wadsworth W~J and Rarity J~G 2007 {\em
  Phys. Rev. Lett.\/} {\bf 99} 120501
  \urlprefix\url{http://link.aps.org/doi/10.1103/PhysRevLett.99.120501}

\bibitem{Martin2012b}
Martin A, Alibart O, {De Micheli} M~P, Ostrowsky D~B and Tanzilli S 2012 {\em
  New J. Phys.\/} {\bf 14} 025002
  \urlprefix\url{http://stacks.iop.org/1367-2630/14/i=2/a=025002?key=crossref.a3b48db56b769594558892412d9c5971}

\bibitem{Bruno2013}
Bruno N, Martin A and Thew R~T 2013 {\em arXiv:1309.6172\/}  5
  (\textit{Preprint} \eprint{1309.6172})
  \urlprefix\url{http://arxiv.org/abs/1309.6172}

\bibitem{Mosley2008}
Mosley P~J, Lundeen J~S, Smith B~J, Wasylczyk P, U'Ren A~B, Silberhorn C and
  Walmsley I~A 2008 {\em Phys. Rev. Lett.\/} {\bf 100} 133601
  \urlprefix\url{http://link.aps.org/doi/10.1103/PhysRevLett.100.133601
  http://prl.aps.org/pdf/PRL/v100/i13/e133601}

\bibitem{Mosley_conditional_2008}
Mosley P~J, Lundeen J~S, Smith B~J and Walmsley I~A 2008 {\em New J. Phys.\/}
  {\bf 10} 093011
  \urlprefix\url{http://iopscience.iop.org/1367-2630/10/9/093011}

\bibitem{Spring2013}
Spring J~B, Salter P~S, Metcalf B~J, Humphreys P~C, Moore M, Thomas-Peter N,
  Barbieri M, Jin X~M, Langford N~K, Kolthammer W~S, Booth M~J and Walmsley I~A
  2013 {\em Opt. Express\/} {\bf 21} 13522
  \urlprefix\url{http://www.opticsinfobase.org/abstract.cfm?URI=oe-21-11-13522}

\bibitem{Harder2013}
Harder G, Ansari V, Brecht B, Dirmeier T, Marquardt C and Silberhorn C 2013
  {\em Opt. Express\/} {\bf 21} 13975
  \urlprefix\url{http://www.opticsinfobase.org/abstract.cfm?URI=oe-21-12-13975}

\bibitem{Soller2011}
S\"{o}ller C, Cohen O, Smith B~J, Walmsley I~A and Silberhorn C 2011 {\em Phys.
  Rev. A\/} {\bf 83} 1--4
  \urlprefix\url{http://link.aps.org/doi/10.1103/PhysRevA.83.031806}

\bibitem{Evans2010}
Evans P, Bennink R, Grice W, Humble T and Schaake J 2010 {\em Phys. Rev.
  Lett.\/} {\bf 105} 253601

\bibitem{Yabuno2012}
Yabuno M, Shimizu R, Mitsumori Y, Kosaka H and Edamatsu K 2012 {\em Phys. Rev.
  A\/} {\bf 86} 010302

\bibitem{Jin2013}
Jin R~B, Shimizu R, Wakui K, Benichi H and Sasaki M 2013 {\em Opt. Express\/}
  {\bf 21} 10659 (\textit{Preprint} \eprint{arXiv:1303.2778})
  \urlprefix\url{http://www.opticsinfobase.org/abstract.cfm?URI=oe-21-9-10659}

\bibitem{Ikuta2013}
Ikuta R, Kobayashi T, Kato H, Miki S, Yamashita T, Terai H, Fujiwara M,
  Yamamoto T, Koashi M, Sasaki M, Wang Z and Imoto N 2013 {\em Phys. Rev. A\/}
  {\bf 88} 042317 (\textit{Preprint} \eprint{1304.0304})
  \urlprefix\url{http://link.aps.org/doi/10.1103/PhysRevA.88.042317
  http://arxiv.org/abs/1304.0304}

\bibitem{Guerreiro2013}
Guerreiro T, Martin A, Sanguinetti B, Bruno N, Zbinden H and Thew R~T 2013 {\em
  Opt. Express\/} {\bf 21} 27641--27651

\bibitem{Fejer1992}
Fejer M, Magel G, Jundt D~H and Byer R 1992 {\em IEEE J. Quantum Electron.\/}
  {\bf 28} 2631--2654

\bibitem{Ljunggren2005}
Ljunggren D and Tengner M 2005 {\em Phys. Rev. A\/} {\bf 72} 062301
  \urlprefix\url{http://link.aps.org/doi/10.1103/PhysRevA.72.062301}

\bibitem{Osorio2013}
Osorio C~I, Sangouard N and Thew R~T 2013 {\em J. Phys. B: At. Mol. Opt.
  Phys\/} {\bf 46} 055501

\bibitem{BROWN1956}
{Hanbury Brown} R and Twiss R~Q 1956 {\em Nature\/} {\bf 177} 27--29
  \urlprefix\url{http://www.nature.com/doifinder/10.1038/177027a0}

\bibitem{Tapster1998}
Tapster P~R and Rarity J~G 1998 {\em J. Mod. Opt.\/} {\bf 45} 595--604
  \urlprefix\url{http://www.tandfonline.com/doi/abs/10.1080/09500349808231917}

\bibitem{Christ2011}
Christ A, Laiho K, Eckstein A, Cassemiro K~N and Silberhorn C 2011 {\em New J.
  Phys.\/} {\bf 13} 033027

\bibitem{Sekatski2012}
Sekatski P, Sangouard N, Bussi\`{e}res F, Clausen C, Gisin N and Zbinden H 2012
  {\em J. Phys. B: At., Mol. Opt. Phys.\/} {\bf 45} 124016
  \urlprefix\url{http://stacks.iop.org/0953-4075/45/i=12/a=124016?key=crossref.4f825b9cc7a3d4c29c8b082dacb007fd}

\end{thebibliography}

\end{document}